\documentclass[aip, amsmath, amssymb, reprint]{revtex4-2}
\usepackage{graphicx}
\usepackage{hyperref}
\usepackage{dcolumn}
\usepackage[utf8]{inputenc}
\usepackage[T1]{fontenc}
\usepackage{physics}
\usepackage{tabularx}

\makeatletter
\def\@email#1#2{
 \endgroup
 \patchcmd{\titleblock@produce}
  {\frontmatter@RRAPformat}
  {\frontmatter@RRAPformat{\produce@RRAP{*#1\href{mailto:#2}{#2}}}\frontmatter@RRAPformat}
  {}{}
}
\makeatother

\begin{document}
\preprint{AIP/123-QED}
\title{Use of Non-Maximal entangled state for free space BBM92 quantum key distribution protocol}

\author{Ayan Biswas}
\affiliation{Quantum Science and Technology Laboratory, Physical Research Laboratory, Ahmedabad, 380009, Gujarat, India.}
\affiliation{York Centre for Quantum Technologies, School of Physics Engineering and Technology, University of York, Heslington, YO10 5DD, York, UK}

\author{Sarika Mishra}
\affiliation{Quantum Science and Technology Laboratory, Physical Research Laboratory, Ahmedabad, 380009, Gujarat, India.}
\affiliation{Indian Institute of Technology, Gandhinagar, 382355, Gujarat, India}

\author{Satyajeet Patil}
\affiliation{Quantum Science and Technology Laboratory, Physical Research Laboratory, Ahmedabad, 380009, Gujarat, India.}
\affiliation{Indian Institute of Technology, Gandhinagar, 382355, Gujarat, India}

\author{Anindya Banerji}
\affiliation{Centre for Quantum Technologies, National University of Singapore, 117543, Singapore}

\author{Shashi Prabhakar}
\affiliation{Quantum Science and Technology Laboratory, Physical Research Laboratory, Ahmedabad, 380009, Gujarat, India.}

\author{Ravindra P. Singh}
\affiliation{Quantum Science and Technology Laboratory, Physical Research Laboratory, Ahmedabad, 380009, Gujarat, India.}

\date{\today}

\begin{abstract}
Satellite-based quantum communication for secure key distribution is becoming a more demanding field of research due to its unbreakable security. Prepare and measure protocols such as BB84 consider the satellite as a trusted device, fraught with danger looking at the current trend for satellite-based optical communication. Therefore, entanglement-based protocols must be preferred since, along with overcoming the distance limitation, one can consider the satellite as an untrusted device too. E91 protocol is a good candidate for satellite-based quantum communication; but the key rate is low as most of the measured qubits are utilized to verify a Bell-CHSH inequality to ensure security against Eve. An entanglement-based protocol requires a maximally entangled state for more secure key distribution. The current work discusses the effect of non-maximality on secure key distribution. It establishes a lower bound on the non-maximality condition below which no secure key can be extracted. BBM92 protocol will be more beneficial for key distribution as we found a linear connection between the extent of violation for Bell-CHSH inequality and the quantum bit error rate for a given setup.
\end{abstract}


\maketitle

\section{Introduction}\label{sec:intro}
In classical communication, the security of encryption keys for parties communicating with each other depends upon the hardness of breaking the encryption algorithm \cite{RevModPhys.74.145, S0097539795293172}. This security is insufficient to protect encrypted messages sent through a public channel once a quantum computer intercepts them. Therefore, with advancements in the development of practical quantum computers, the demand for information-theoretic secure communication based on the principles of physics has increased. It has already been demonstrated that using Shor's quantum algorithm, one can break most of the encryption techniques applied in classical key distribution between communicating parties, say Alice \& Bob \cite{S0097539795293172, RevModPhys.74.145, BENNETT20147}. Quantum key distribution (QKD) uses the principles of quantum mechanics to securely distribute keys between the two communicating parties \cite{BENNETT20147, PhysRevLett.67.661}. Moreover, using QKD also ensures that Eavesdropper's presence can be detected in real-time by observing the disturbance in the channel, unlike conventional classical key distribution \cite{shor2000simple, lucamarini2018overcoming, gottesman2004security}.

Based on the usage and type of encryption, several QKD protocols are available, e.g., BB84 \cite{BENNETT20147, RevModPhys.74.145}, SARG04 \cite{scarani2004quantum}, COW \cite{stucki2009high}, E91 \cite{PhysRevLett.67.661}, etc. The BB84 protocol is widely applied due to its ease of implementation and the existence of composable security proofs for practical deployments \cite{shor2000simple, Pirandola:20}. However, it is prone to side-channel attacks \cite{lucamarini2012device}, and distance is limited as the disturbance in the channel increases with the propagation. Entanglement-based QKD (EBQKD) protocol can tackle the challenge of distance limitation for secure key transmission as posed by BB84 \cite{PhysRevLett.67.661, PhysRevA.65.052310, brassard2000limitations}. The security of EBQKD comes from the principles of no-cloning and monogamy of quantum entanglement \cite{Pirandola:20, coffman2000distributed, RevModPhys.92.025002}. The latter states that if two parties (Alice \& Bob) share a maximally entangled state, the third party cannot have any correlation with the communicating parties \cite{coffman2000distributed}. EBQKD is also ideal for satellite-based quantum communication by sharing entangled photons between the two ground stations to communicate securely \cite{PhysRevLett.120.030501, villar2020entanglement}. Security in EBQKD is ensured by checking violation of Bell's inequality, which makes the protocol robust against all strategic attacks. Even without checking Bell-CHSH inequality, one can still distribute secret keys if they share a maximally entangled state, like in BBM92 protocol \cite{bennett1992quantum}.

For carrying out long-distance QKD, e.g., satellite-based quantum communication, EBQKD has an advantage as it connects two distantly situated ground stations with a single satellite \cite{10.48550/arxiv.2210.02229}. Considering EBQKD for practical purposes, the BBM92 protocol is less resource intensive than the E91 protocol ensuring the same security using a maximally entangled state. The key rate is higher in the BBM92 protocol as a violation of the Bell-CHSH inequality is not always required in building real-time secret keys. In this article, we find the relation between the quantum bit-error rate (QBER) and Bell-CHSH parameter $S$, including experimental imperfections in field-based QKD experiments. This connection between QBER and $S$ can indicate the purity of the source related to the QBER generated in real-time. A similar process was done only after sacrificing many key bits for checking $S$ separately, then going for secret key extraction by looking to QBER \cite{fujiwara2014modified}. Therefore the present correlation between QBER and $S$ comes in handy in providing a longer secret key for the same raw key. Additionally, we also determined the mutual information (MI) shared between Alice, Bob and Eve, which further introduces a limit on the secret key rate per bit. If one uses a non-maximal entangled state, then the Bell-CHSH inequality can be violated with low detector efficiency, closing all loopholes, and moving closer to the realization of device-independent QKD (DIQKD) systems \cite{fuchs1997optimal, eberhard1993background}.

This work aims to experimentally verify the variation of $S$ with QBER and set a minimum bound on $S$ for the safe operation of the BBM92 protocol. This study contains section \ref{sec:TB} that describes the theoretical background of the present work. The experimental method to generate non-maximally entangled state is elaborated in section \ref{sec:EM}, and the results are presented in section \ref{sec:RnD}. Section \ref{sec:conc} concludes our work with suggestions to implement it in real scenarios, and highlight the applicability of the BBM92 protocol with source imperfections for secure long-distance communication.

\section{Theoretical Background}\label{sec:TB}
\begin{figure}[h]
    \centering
    \includegraphics[width=7.5cm]{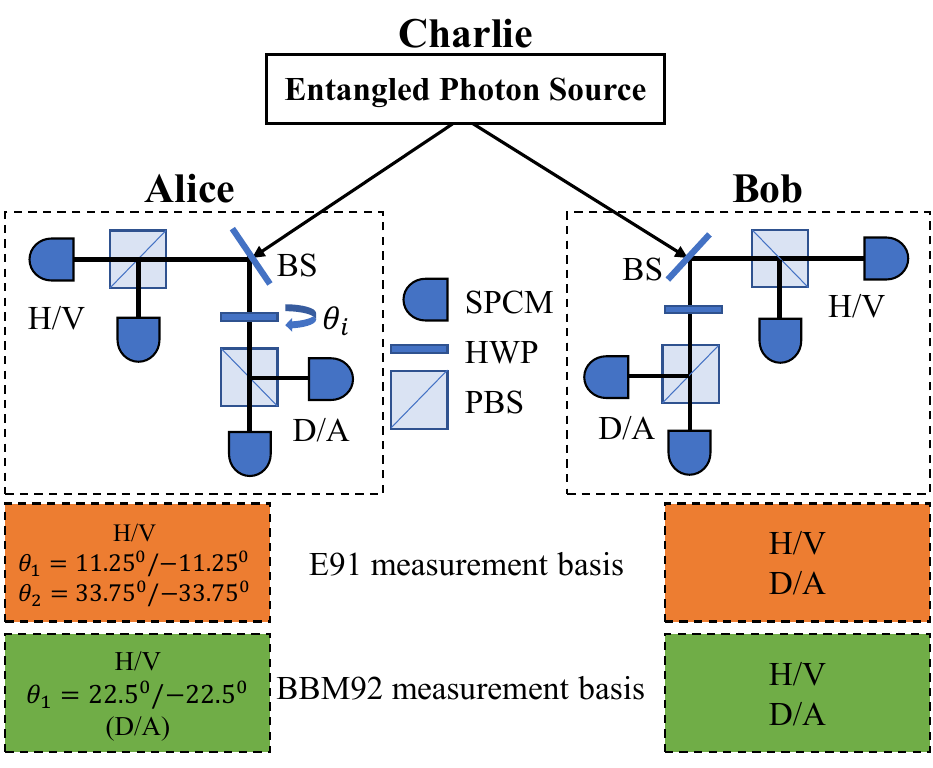}
    \caption{Schematic for EBQKD including basis measurements for Alice and Bob in E91 and BBM92 protocols. The HWP at $\theta_i$ defines the angles for various basis states required for E91 and BBM92 protocol.}
    \label{fig:Protocol}
\end{figure}

In standard EBQKD, as shown in Fig. \ref{fig:Protocol}, a common sender, Charlie, sends a pair of polarization-entangled photons to Alice and Bob through a quantum channel (fiber or free space). Alice and Bob independently make their measurements on chosen random bases. The measurement bases are different for the E91 and BBM92 protocols, also shown in Fig. \ref{fig:Protocol}. After the measurement, Alice and Bob declare their basis choice through the public channel and build the secure key for encryption.

E91 protocol, in principle, is secure against any eavesdropping strategy \cite{coffman2000distributed}. Alice and Bob will only form the key when they choose the same basis for their measurements. The rest of the measurement results will go for calculating the Bell-CHSH parameter $S$ for the protocol's security. Ideally, if a maximally entangled state is used, any value of the Bell-CHSH parameter below $2\sqrt{2}$ will render this protocol insecure. However, the quality of the quantum channel might adversely affect the value of $S$, and implementing this protocol can be challenging as the number of photon pairs may degrade. This poor correlation results in information leakage to Eve, which increases her chances of gaining access to the key. Also, the drawback of this protocol is that it has a low key rate as most of the generated raw bits from the measurements are used for security checks through violation of Bell's inequality. 

In BBM92 protocol, a secret key can be extracted without Bell state analysis if one has a maximally entangled photon pair source \cite{bennett1992quantum}. The protocol is similar to E91, and the difference lies in the measurement bases, which are \{H/V, D/A\} for both Alice and Bob. The key is generated when Alice and Bob measure in compatible bases. The primary advantage of BBM92 over E91 is that the key rate becomes considerably higher as a majority of the detection events are used to build the key, and very few are utilized to check for QBER. The QBER threshold for secure key distribution is the same as that of BB84 protocol \cite{PhysRevA.65.052310}. So, if one has a maximally entangled state, one can perform EBQKD without Bell-CHSH measurement \cite{bennett1992quantum, Pirandola:20, PhysRevLett.59.2044}.

In the BB84 protocol, a QBER ($\delta$) of 11\% can be tolerated against collective attacks as the key rate $r$ goes to zero above that according to the relation $r=(1-2H(\delta))$ \cite{shor2000simple}. The same error rate is also true for the BBM92 protocol \cite{shor2000simple, PhysRevA.65.052310}. Therefore, by looking at the correlation between QBER and $S$, one can interpret the extent of non-maximal entangled photons that can be used for EBQKD. For a perfectly secure QKD protocol, one needs a maximally entangled source to attain the maximum value of $S$. The increased non-maximality of the entangled photon source may leak information to Eve \cite{acin2007device}. This indicates that by entanglement monogamy, Eve can have some correlation either with Alice or Bob \cite{coffman2000distributed, Pirandola:20}. This can also be checked directly with the formula given by \cite{acin2007device, fujiwara2014modified}
\begin{equation}
    I(A:E) = H\left(\frac{1+\sqrt{S^2/4-1}}{2} \right),
    \label{Alice_Eve_MI}
\end{equation}
where $I(A:E)$ is the mutual information (MI) that can be shared between Alice and Eve, $H$ is the binary entropy, and $S$ is the Bell-CHSH parameter. The maximum amount of information shared by Alice and Bob between each other for the BBM92 protocol is $I(A:B)$. This can be calculated using the relation
\begin{equation}
    I(A:B)=H(A)+\sum_{a \epsilon A} p(a) \sum_{b \epsilon B} p(b\mid a) \ Log \ p(b \mid a),
    \label{Alice_Bob_MI1}
\end{equation}
where $p(a)$ is the probability of getting a polarization (say $\ket{H}$) at Alice or Bob out of four polarization states. $p(b\mid a)$ is the probability of getting a polarization ($\ket{V}$) at Bob, given polarization ($\ket{H}$) is measured by Alice or vise versa. In experiments, this quantity can also be calculated by measuring bit-error ($e_b$) and phase-error ($e_p$) in the system. For EBQKD, the mutual information between two parties is given by \cite{Pirandola:20}
\begin{equation}
    I(A:B)=1-H(e_b)-H(e_p).
    \label{Alice_Bob_MI2}
\end{equation}
Experimentally, MI can be calculated from the coincidences detected at both ends normalized by the individual detector counts. The final secure key rate of the protocol can be written as \cite{devetak2005distillation, shor2000simple},
\begin{equation}
    r=I(A:B)-I(A:E).
    \label{Keyrate1}
\end{equation}
where $r$ is the secret key rate per bit. Secure key extraction is possible when $r\geq0$; this implies $I(A:B)>I(A:E)$. Since both these quantities vary with QBER and $S$, obtaining a range for both would enable secure key extraction, efficiently.

\section{Experimental Method}\label{sec:EM}
\begin{figure}[h]
    \centering
    \includegraphics[width=7.5cm]{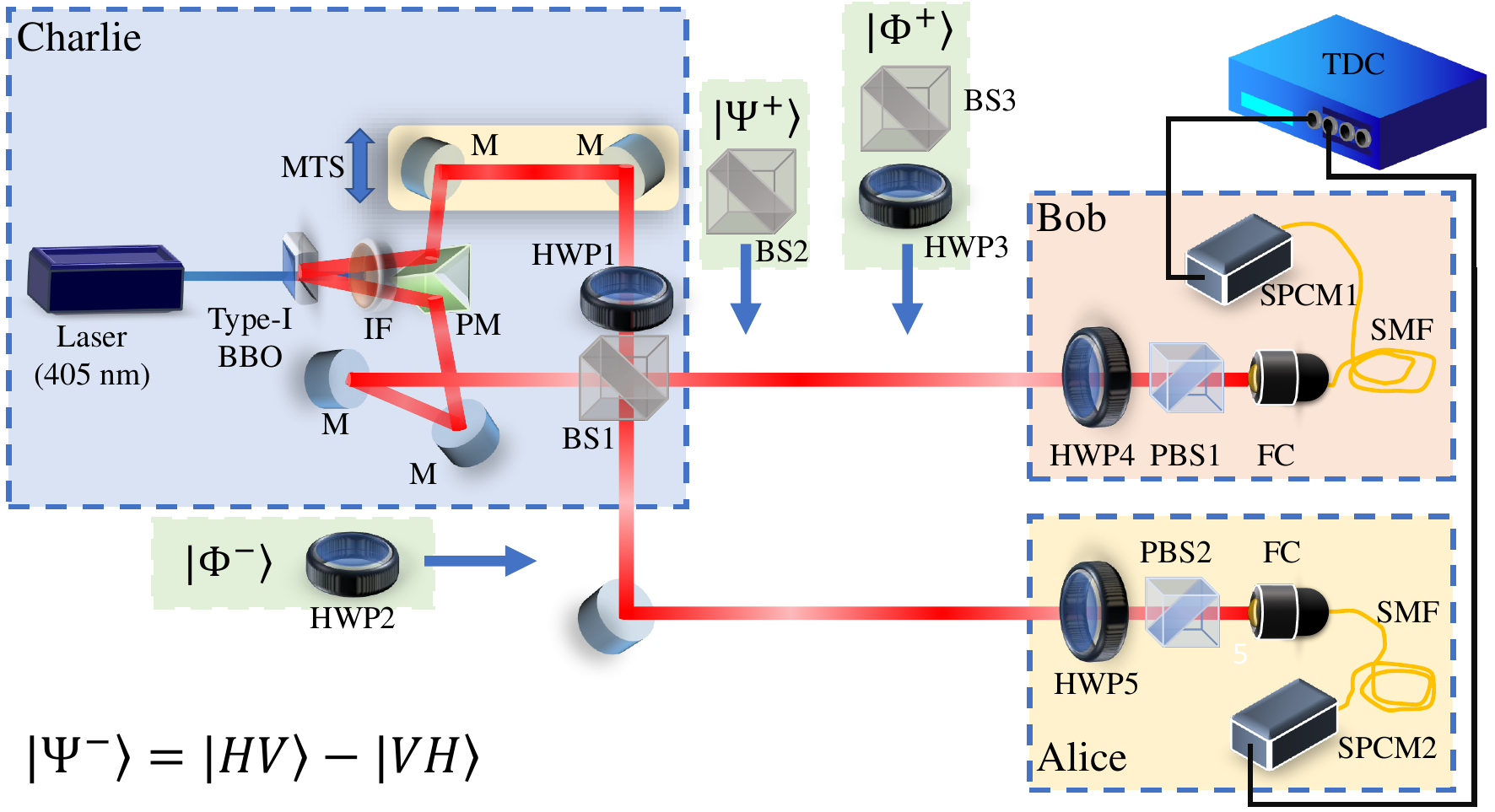} 
    \caption{Experimental scheme for generating $\ket{\Psi^-}$ from HOM interferometer. All the other four Bells states have also been created by introducing appropriate optics. \textbf{BS}: 50:50 Beam-Splitter; \textbf{PBS}: Polarizing Beam Splitter; \textbf{IF}: Interference filter; \textbf{MTS}: Motorized Translation Stage; \textbf{M}: Mirror; \textbf{HWP}: Half Wave Plate; \textbf{PM}: Prism mirror; \textbf{SPCM}: Single Photon Counting Module; \textbf{SMF}: Single Mode Fiber; \textbf{TDC}: Time to Digital-Converter.}
    \label{fig:ExperimentalSetup}
\end{figure}

We have used the Hong-Ou-Mandel interferometer (HOM) technique to generate the desired non-maximally entangled photon state \cite{fuchs1997optimal, bouchard2020two, PhysRevLett.59.2044}. Figure \ref{fig:ExperimentalSetup} shows the schematics for the experimental setup to generate all four Bell states, and the advantage is that their maximality is controlled by controlling the HOM visibility. A laser of wavelength 405 nm pumps a nonlinear crystal (Type-I BBO) to produce degenerate photon pairs by the nonlinear spontaneous parametric down-conversion (SPDC) process. A prism mirror (PM) separates the pathways of two generated photons. The HOM interference resulting in photon bunching will only occur when indistinguishable photon pairs overlap, indicating a coincidence dip at the detector. There can be multi-photon pairs (less probable) coming out of the SPDC process that can increase QBER; however these are filtered out by HOM interferometer. To obtain the desired entangled state, the polarization of one of the two photons is changed by placing an HWP1 in one of the arms after the prism mirror.

\begin{figure}[h]
    \centering
    \includegraphics[width=7.5cm]{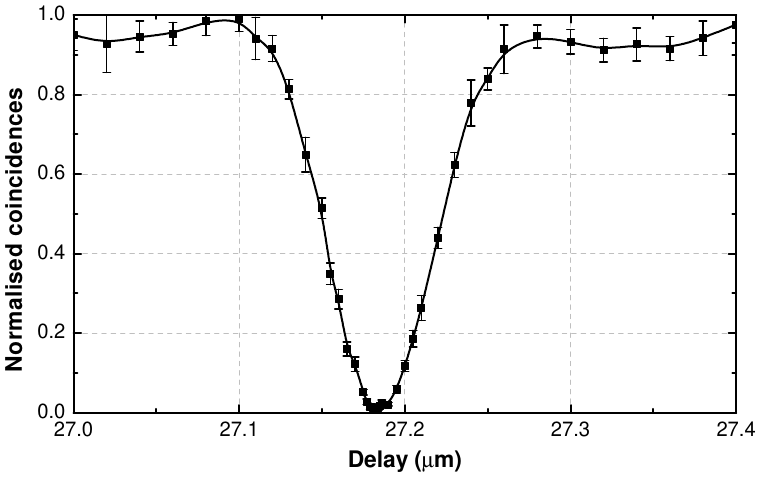}
    \caption{HOM dip is obtained when both the photons are completely indistinguishable. The delay is introduced between the two input ports to introduce the temporal distinguishability.}
    \label{fig:HOM}
\end{figure}

The experimentally observed visibility of the HOM dip by controlling the motorized translation stage (MTS) is shown in Fig. \ref{fig:HOM}. At the HOM dip region, if one of the incoming arms is changed to orthogonal polarization (HWP1), then we have two distinguishable photons falling at the BS1, resulting to four possibilities, and the output state can be written as,
\begin{equation}
    \label{BS:out}
    \ket{\Psi}_{out}\propto \left( \alpha_1\ket{H_1V_1} + \alpha_2\ket{H_1V_2} +\alpha_3 \ket{H_2V_1} +\alpha_4 \ket{H_2V_2} \right),
\end{equation}
where $\alpha_i$ are the complex amplitudes of the corresponding state. After post-selecting the simultaneously detected photon pairs at the output port of the BS1, the above state will become an entangled state. Table \ref{tab:StateGeneration} summarizes the settings to obtain the desired non-maximally entangled states
\begin{equation}
    \ket{\Psi}_{int}\propto \left(\alpha_2\ket{H_1V_2} +\alpha_3 \ket{H_2V_1} \right).
    \label{eq7}
\end{equation}

\begin{table}[h]
    \centering
    \caption{Arrangement of optics required for the Bell states generation using HOM.}
    \label{tab:StateGeneration}
    \begin{tabular}{|l|p{2.5cm}|p{5cm}|}
        \hline
        State & Optics & Position \\
        \hline
        $\ket{\Psi^-}$ & HWP1 $@45^{\circ}$ & Before BS1 \\ 
        \hline
        $\ket{\Phi^-}$ & HWP2 $@45^{\circ}$ & At any of the output port of BS1 \\
        \hline
        $\ket{\Psi^+}$ & BS2 & After the BS1 (in any arm either reflected or transmitted). Only output port of BS2 will be used to observe entanglement \\
        \hline
        $\ket{\Phi^+}$ & BS3 and HWP3 $@45^{\circ}$ & BS3  at any of the output port of BS1 and HWP3 at either of the output port of the BS3. Entanglement can be observed at the output ports of BS3 \\
        \hline
     \end{tabular}
\end{table}

The generation of $\ket{\Psi^-}$ state is shown in Fig. \ref{fig:ExperimentalSetup}, and the rest of the Bell states can be obtained by using the appropriate optics as shown in Table \ref{tab:StateGeneration}. All the states are then measured by projecting them to different polarization states using a combination of HWP and PBS, which is then detected through single-mode fiber-coupled single photon counting modules (SPCM). These can be thought of as the detection setup for Alice and Bob. The coincidences from both detectors are recorded for various polarization projections (by rotating the HWP4 and HWP5), typically used in the BBM92 protocol. Coincidences in the same basis for the state $\ket{\Phi^{\pm}}$ will give the key rate estimation for the BBM92 protocol. While for the state $\ket{\Psi^{\pm}}$, coincidences on a complimentary basis will form a sifted key (anti-correlated photon polarizations will form key as the state is $\ket{HV}\pm\ket{VH}$).

\section{Results and Discussion}\label{sec:RnD}
The measurement of QBERs is performed from the coincidence counts by varying the HOM visibility. These QBER results can then be used in the Eq. \ref{Alice_Bob_MI2} to calculate the MI ($I(A:B)$) between Alice and Bob for corresponding Bell states. Coincidence counts for all the specific combinations of polarization are recorded by adjusting HWPs angle (HWP4 and HWP5) to calculate Bell-CHSH parameter ($S$) and key-rate estimation. We measured the Bell-CHSH parameter for each of the four Bell states with different visibility settings. This visibility in HOM will change the coefficients of the corresponding states generated for EBQKD. We record the coincidences for key rate estimation with the change in the amount of entanglement (i.e., change in $\alpha_{i},S$). This will indicate the variation of $S$ with QBER.

This study experimentally proves the relationship between two important parameters in QKD protocols, QBER and $S$. Our experimental results have shown a linear relationship between QBER and $S$, with a negative slope, which is valid for individual attacks and is given by \cite{fuchs1997optimal},
\begin{equation}\label{eq:svsqber}
    S=2 \sqrt{2}(1-2 \delta).
\end{equation}
where $\delta$ is the disturbance in the signal. The error limit for the QKD protocol can be determined from the value of $S$, as it indicates the strength of the correlation between Alice and Bob's measurements. If the value of $S=2$, then $\delta=14\%$, interpreting that if the QBER is higher than 14$\%$, Eve could potentially have knowledge of the key. For collective attacks, the error limit is lower, at 11$\%$. For a given channel, this relation is helpful as it directly connects $S$ with QBER. Specific QBER received by Alice or Bob can directly indicate the value of $S$ for that particular system. This can be a double check in the security if one is doing EBQKD without sacrificing extra bits for the Bell test.

\begin{figure}
   \centering
   \includegraphics[width=7.5cm]{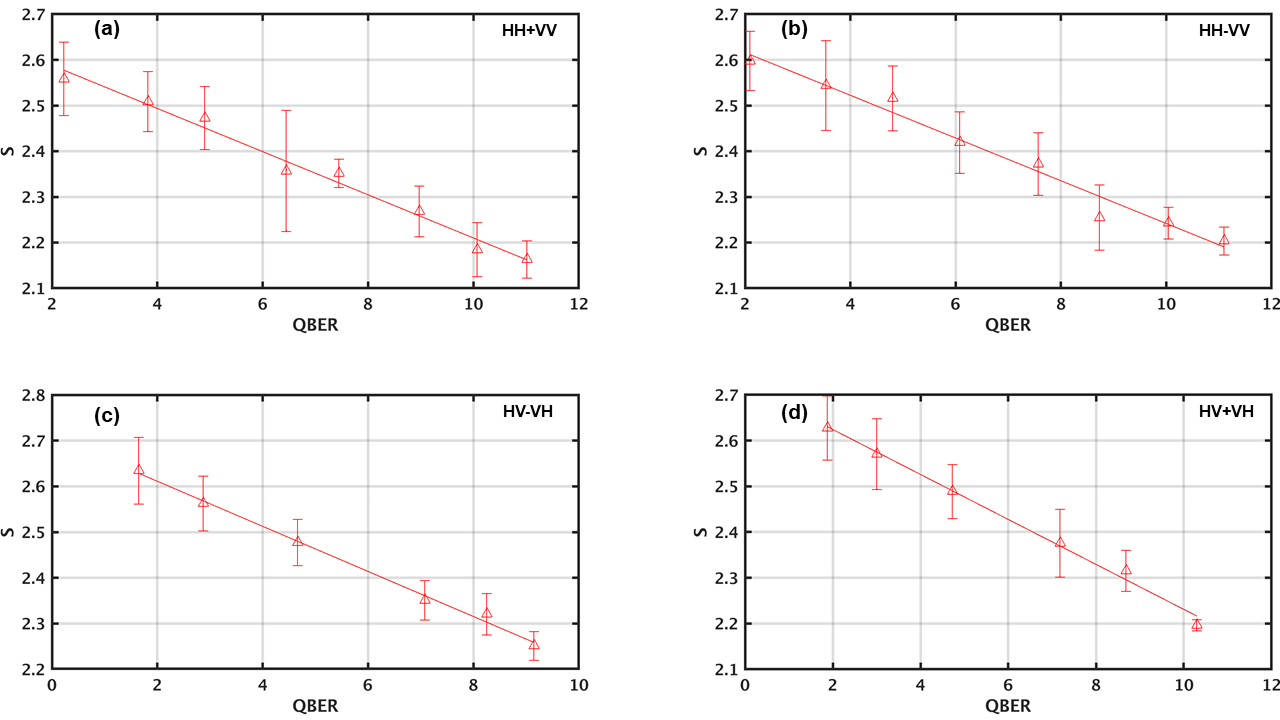}
   \caption{Variation of Bell's inequality ($S$) with QBER (\%) for all four Bell states; a) $\ket{\phi^+}$, b) $\ket{\phi^-}$, c) $\ket{\psi^-}$, and d) $\ket{\psi^+}$. The error bar in QBER is small enough to be observed on the plot.}
   \label{fig:SvsQBER}
\end{figure}

The figure \ref{fig:SvsQBER}(a) shows the variations of $S$ with QBER for state $\ket{\phi^+}=C_1\ket{H_1H_2}+C_2\ket{V_2V_1}$. The maximum recorded value of Bell's inequality parameter is $2.64\pm0.12$ for which the QBER is 2\%.  The graph matches well with the predicted value of the error bound of the BB84 protocol. The minimum value of the Bell parameter to run the protocol safely is 2.1. This indicates that the amount of non-maximality that can be achieved is 2.1 for secure key distribution in the BBM92 protocol. Similarly for another Bell states, the variation of $S$ with QBER for $\ket{\phi^-}=C_1\ket{H_1H_2}-C_2 \ket{V_2V_1}$, $\ket{\Psi^+}=C_1\ket{H_1V_2}+C_2\ket{H_2V_1}$ and $\ket{\Psi^-}=C_1\ket{H_1V_2}-C_2\ket{H_2V_1}$ are shown in Fig. \ref{fig:SvsQBER}(b-d), respectively. Irrespective of any Bell state, the BBM92 protocol results in the same error bound as the BB84 protocol, including implementation discrepancies. This will not affect the variation of $S$ with QBER for a given system in the protocol. 

The presented experimental results in the Fig. \ref{fig:SvsQBER} are in good agreement with the theory. The experiment assumes identical detector efficiency for Alice and Bob, whereas the overall transmission efficiency could vary due to different channel lengths. The Fig. \ref{fig:SvsQBER} illustrates the effect of changing QBER on the value of $S$, which is essential to determine the condition of the source in the transmitting end. The agreement of the relationship for all four types of Bell states confirms the robustness of the results under experimental discrepancies. Also, having entanglement non-necessarily gives a secure key. Eve might get the advantage in gaining the information from the weakness in the entanglement of the source. This will further reduce the bound in error to extract the secret key.

\begin{figure}
    \centering
    \includegraphics[width=7.5cm]{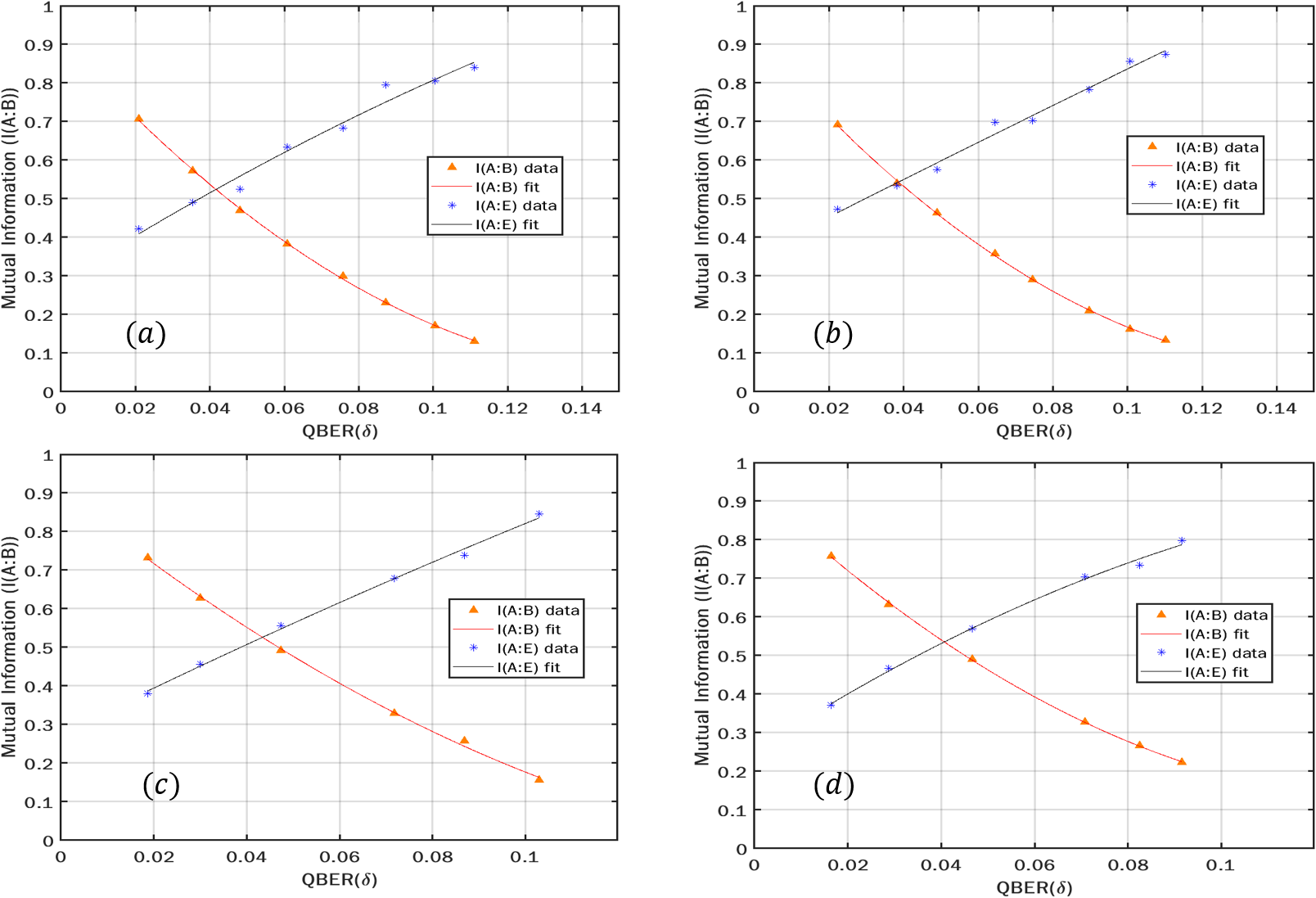}
    \caption{Variation of Mutual Information between Alice-Bob ($I(A:B)$) and Alice-Eve ($I(A:E)$) with QBER for all four Bell states; a) $\ket{\phi^+}$, b) $\ket{\phi^-}$, c) $\ket{\psi^-}$, and d) $\ket{\psi^+}$.}
    \label{fig:MIvsQBER}
\end{figure}
For calculating the secure key rate, the difference between the mutual information of Alice-Bob ($I(A:B)$) and Alice-Eve ($I(A:E)$) is considered. The key rate can be calculated using Eq. \ref{Keyrate1}. The plots for MI between Alice Bob and Alice Eve are shown in Fig. \ref{fig:MIvsQBER}. The plots show that non-zero secure key rates are only possible for error bounds up to $\sim4\%$, obtained for $I(A:B)>I(A:E)$. Above this, even though one has entanglement but still the secure key rate extraction won't be possible. The attack strategy by Eve is taken to be general as she uses the weakness in entanglement to gain information about the key. In Eq. \ref{Keyrate1} for key rate $r$, it is assumed that Eve can perform any kind of attack, and have advantage as Alice and Bob are not using non-maximal entanglement. The information leakage is because the states in the QKD are not perfectly entangled. Figure \ref{fig:MIvsQBER} shows the secret key rate for the four Bell states in experimental conditions. 

By understanding this relationship, researchers can generate longer secret keys from satellite-based systems with shorter pass times without sacrificing too much of their raw key material to perform the necessary tests. This is important because the quality of the entangled photon source can degrade over time, and it may not be possible to maintain a maximally entangled state. One can use a non-maximally entangled state for QKD, provided they have already calibrated the source, and the leakage due to error is also taken into account. The discrepancies in the source will decide the intrinsic error, that needs to be added on the top of the QBER while distilling the keys. To make the protocol secure against Eve, one has to consider this error, apart from the QBER. This will make the key generation process less resilient against errors in channel than the expected one (because one has to consider the QBER due to  non-maximal entangled source). Due to non-maximality of the source, Eve can extract some amount of information, and this leaked information can be removed while distilling the keys. This has to be done even if one is observing a Bell-CHSH violation.

\section{Conclusion}\label{sec:conc}
This study highlights connection between the violation of the Bell-CHSH inequality and QBER in QKD protocols. The relationship between $S$ and QBER is independent of the Bell state used in the protocol, and can be used to extract secret keys safely for the BBM92 protocol, even if the source is not maximally entangled. The knowledge of this relationship enables the estimation of $S$ from the QBER, which allows for error correction and privacy amplification accordingly. This connection directly indicates whether or not the quantum channel is being tampered. Importantly, the value of $S$ can be calibrated with the corresponding QBER value before the QKD protocol is initiated. This calibration ensures that the error limit is set appropriately, and that the secret key rate generated by Alice and Bob is maximized. This calibration is particularly important in satellite payloads, as it ensures that the QKD protocol is robust and reliable even in the harsh space conditions.

In the present work, we also studied additional bound on QBER arising from the mutual information shared between Alice and Eve. Using non-maximal entangled states in QKD can be more beneficial for long-distance communication as they are more robust against source and channel disturbances. Also, maintaining the entangled photon source becomes easier, as maximality is not always required. The present study can help to do long-term QKD without routine system characterizations. The current work finds application in satellite-based QKD or free space QKD over a long time without further characterizations at each run.

\section*{Acknowledgments}
The authors like to acknowledge the funding support from the Department of Science and Technology (DST), India through QuEST program.

\section*{Disclosures}
The authors declare no conflicts of interest related to this article.

\section*{References}
\bibliography{manuscript}

\end{document}